\documentclass[10pt]{article}
\usepackage{geometry}
\usepackage{enumerate}
\usepackage{latexsym,booktabs}
\usepackage{amsmath,amssymb}
\usepackage{graphicx}
\usepackage{color}
\usepackage{amsmath}
\usepackage{scalefnt}
\usepackage{hyperref}
\usepackage{amsthm}
\usepackage{mathrsfs}
\usepackage{multirow}
\usepackage{fancyvrb}
\usepackage{comment}
\usepackage{framed}
\usepackage{caption}
\usepackage{multicol}
\usepackage{multirow}
\usepackage{eurosym}
\usepackage{amsopn}
\usepackage{amssymb}
\usepackage[dvipsnames]{xcolor}
\usepackage{array}
\usepackage{amsthm}
\usepackage[toc,page]{appendix}
\usepackage{wrapfig}
\usepackage{indentfirst}
\usepackage{physics}
\usepackage{rotating}
\usepackage{apalike}
\usepackage{color}
\usepackage{listings}
\usepackage{caption}
\usepackage{subcaption}
\usepackage{soul}
\usepackage{xcolor}

\numberwithin{Theorem}{section}
\numberwithin{Definition}{section}
\numberwithin{Lemma}{section}
\numberwithin{Algorithm}{section}
\numberwithin{equation}{section}

\definecolor{gray}{rgb}{0.5,0.5,0.5}

\DeclareFontShape{OT1}{cmtt}{bx}{n}{<5><6><7><8><9><10><10.95><12><14.4><17.28><20.74><24.88>cmttb10}{}

\title{Point process models for quasi-periodic volcanic earthquakes}
\author{Anastasia Ignatieva, Andrew F. Bell, Bruce J. Worton \\ \emph{University of Edinburgh}}

\begin{document}

\maketitle

\begin{abstract}
Low frequency or long period (LP) earthquakes are a common phenomenon at active volcanoes, and are ubiquitous at persistently active andesitic and dacitic subduction zone volcanoes. At these systems, LP earthquakes provide critical information regarding the state of volcanic unrest, and their occurrence rates are key data on which eruption forecasts are based. Point process modelling of volcanic earthquake occurrence allows potential insights into the underlying physical processes driving unrest, and quantitative, probabilistic, eruption forecasts, for example, through application of the Failure Forecast Method (FFM). However, unlike high-frequency volcano-tectonic (VT) earthquakes, which are typically random or clustered in time, LPs are more commonly quasi-periodic or `anti-clustered'. Consequently, the existing Poisson point process methods used to model occurrence rates of VT earthquakes are unlikely to be optimal for LP data. Here we evaluate the performance of candidate inhomogeneous point process formulations of the FFM for quasi-periodic LP data, based on four different inter-event time distributions: exponential (for Poisson), Gamma, inverse Gaussian, and Weibull. Using example LP data recorded before a large explosion at Tungurahua volcano, Ecuador, we examine how well these models explain the observed data, and the quality of retrospective forecasts of eruption time. We use a Markov chain Monte Carlo approach to estimate parameter posterior distributions within a Bayesian framework. Goodness-of-fit is assessed using Quantile-Quantile and Kolmogorov-Smirnov methods, and results are benchmarked against those obtained from idealised synthetic datasets. Inverse Gaussian and Gamma models were both found to fit the data well, with the inverse Gaussian model slightly outperforming the Gamma model. However, retrospective forecasting analysis shows that the Gamma model performs best, with the initial preference for the inverse Gaussian model controlled by catalogue incompleteness late in the sequence. The Gamma model fits the data significantly better than the Poisson model, and simulations show it produces better forecasts for highly periodic data. Simulations also show that forecast precision increases with the degree of periodicity of the earthquake process using the Gamma model, and so should be better for LP earthquakes than VTs. These results provide a new framework for point process modelling of volcanic earthquake time series, and verification of alternative models. 
\end{abstract}

\section{Introduction}

Earthquakes are an important source of information about processes occurring at active volcanoes. Attributes such as the rate \cite{Voight1988}, location \cite{White2016}, and nature \cite{Roman2006,chouet} of earthquakes are key indicators of the state of volcanic unrest, and they form the basis for forecasts of future eruptive activity \cite{Sparks2003}. Changes in the rate of earthquakes are often observed before the onset of eruptions \cite{Voight1988,Kilburn2003a}, or changes in the style and intensity of ongoing eruptions \cite{Salvage2016a,Hotovec2013,bell2}. The ability to retrospectively identify and quantify earthquake rate changes is important to develop an empirical and statistical understanding of pre-eruptive signals \cite{Bell2013}, and to develop conceptual models of the physical process underpinning them. Such analyses provide information that forms the basis of eruption forecasts and early warnings. However, quantifying trends in earthquake time series data can be a challenging problem. This challenge can be particularly acute at volcanoes, where activity can be highly variable, involving a range of different earthquake types, and a range of different temporal statistics.

\subsection{Statistical properties of volcanic earthquakes}
The temporal occurrence of earthquakes can be understood and modelled within the context of a point process \cite{daley}. This approach is widely adopted within the tectonic earthquake and seismic hazard community, where the temporal statistics of earthquake occurrence have been extensively studied \cite{ogata1999}. Point process modelling allows characterisation of changes in the underlying rate of earthquakes, and the nature of interactions between them. Earthquake statistics are less well known in volcanic regions, partly because the nature of volcanic earthquake processes can be far more varied than for tectonic seismicity \cite{chouet,Chouet2013a}. At volcanoes reawakening after extended repose periods, or in tectonically active areas, seismicity is often dominated by high-frequency volcano-tectonic (VT) earthquakes \cite{Kilburn2003a}. These events are usually associated with brittle fracturing or slip along existing fault planes, in response to stress changes associated with magma movement and pressurisation. VT earthquakes share many properties with tectonic earthquakes, and in many cases appear to follow similar empirical relationships, such as the Gutenberg-Richter relation \cite{Roberts2015}, and the modified Omori law, resulting in Poisson or clustered inter-event times \cite{Bell2013}. However, the highly variable stressing rates associated with magma pressurisation and movement means that the background rate can vary quickly, and it can be difficult to isolate from changes in the properties of earthquake interactions.

In contrast, at frequently or persistently active `open-system' volcanoes, particularly with intermediate composition magmas, seismicity is more commonly dominated by low-frequency (or long period; LP) earthquakes. LP earthquakes typically have energy concentrated between 1-5 Hz, emergent onsets and long harmonic coda \cite{chouet}. Changes in the rate of LP earthquakes are often observed in the hours or days before changes in the nature of eruptive activity. Models for LP source processes typically involve two components; an initial excitation or trigger mechanism, and a subsequent modification of the waveform by resonance or scattering processes. Excitation mechanisms that have been proposed include degassing \cite{cruzchouet,molina2004}, hydrothermal fluid movement \cite{matoza2015}, magma flow \cite{julian1994}, and brittle failure of magma \cite{neuberg2006}. These initial processes excite resonances within a fluid-filled crack \cite{chouet} or the magma column \cite{Neuberg2000} resulting in the long, decaying, harmonic coda. A ``dry" LP earthquake mechanism has also been suggested, involving slow rupture of the shallow edifice at low confining pressure, followed by a strong scattering effect \cite{Bean2013a}. Knowledge of the statistical properties of LP earthquakes is limited, but reported observations often include a locally restricted range of inter-event times, associated with quasi-periodic (anti-clustered) behaviour \cite{Powell2003,Bell2017}. Extreme, highly-periodic instances of this behaviour have been reported at several persistently active volcanoes, and have been referred to as ``drumbeat" earthquakes \cite{Bell2017,Iverson2006,Neuberg2000}. The different statistical properties of such quasi-periodic point processes means that their analysis requires a different analytical approach to Poisson or clustered processes \cite{bell2}.

\subsection{Failure forecast method for eruption forecasting} \label{FFM}

Increasing rates of earthquakes have been reported before from a range of different eruption types and sizes, and interpreted in terms of the progressive failure of all or part of the volcanic edifice due to elevated stresses \cite{Voight1988,Kilburn2003a}. Voight (1988) proposed a predictive relation between the rate and acceleration in precursory geophysical data (e.g. seismicity) during the build-up to eruptions, based on both empirical observations and theoretical considerations of the physics of material failure. This approach has become known as the failure forecast method (FFM), and has been widely applied to pre-eruptive earthquake data, commonly in retrospective analyses. Under typical conditions, the FFM expects a power-law increase in the rate of earthquakes with time as the eruption nears \cite{Voight1988}:
\begin{equation}\label{lambda}
\lambda (t) = k (t_f - t)^{-p}
\end{equation}
where $k$ is a constant (related to the amplitude of the signal), $t_f$ is the time of eruption, and $p=\frac{1}{a-1}$ is a parameter describing the non-linearity of acceleration \cite{Bell2013}. At time $t_f$, the rate becomes instantaneously infinite, commonly interpreted as the onset of the eruption process \cite{Voight1988}. 

Although the origins of the FFM are empirical, the physical underpinning of the model is often discussed in terms of progressive material failure of part or all of the volcanic edifice. This interpretation might seem reasonable for sequences dominated by volcano-tectonic earthquakes, with likely source mechanisms involving brittle fracture or stick-slip of faults within the edifice. Each earthquake could be considered as an increment of brittle failure. However, accelerating sequences of LP earthquakes have also been reported before eruptions. Most LP source models invoke a fluid phase either during excitation or resonance, many involve repeated activation of the same or limited number of sources, and non-Poissonian occurrence statistics. Although the evolution of mean earthquake rate sequences are empirically consistent with the FFM, it is less clear how they can be explained by a material failure model. This remains an outstanding question, and one which improved statistical quantification of such sequences might provide new insights.

Different statistical methods have been used to apply the FFM to earthquake data before eruptions. Voight (1988) used an inverse rate linearisation with least squares to estimate model parameters, including eruption time. Bell et al. (2011) investigated accelerating rates of volcano-tectonic earthquakes and laboratory acoustic emissions, and suggested that the error structure of the data was better explained by a Poisson distribution, and used a generalised linear model to apply the FFM to binned rate data \cite{Bell2011}. Bell et al. (2013) develop a maximum-likelihood approach to FFM parameter estimation for VT data, based on an inhomogeneous Poisson process. Bou\'e et al. (2015) applied a Bayesian methodology for real-time estimation of the FFM parameters, finding a Gaussian distribution provided a better model for the event rates in their study. For quasi-periodic earthquakes, it is likely that methods based on inhomogeneous Poisson processes, or linearised least-squares, will not accurately represent the error structure of the data, leading to inaccurate parameter estimates and erroneous forecasts. To accommodate this, Bell et al. (2018) proposed a point process model for quasi-periodic LP earthquakes, using the gamma distribution to describe inter-event times.

\subsection{Point processes and Bayesian methods}

Point processes are defined by a time series of discrete stochastic events or `spikes', occurring in continuous time \cite{daley}. The events are binary, in the sense that at any one time, the point process can only take one of two values, corresponding to whether an event has or has not occurred at that time \cite{chapter2}. For a `homogeneous' Poisson process, the event rates follow a Poisson distribution with constant rate, and the inter-spike intervals (ISIs) follow an exponential distribution with constant rate. The number of spikes in the time interval $[s, s+r]$ is identically distributed for all $s$, and the number of events in non-overlapping intervals is independent. For an `inhomogeneous' Poisson point process, the event rate can vary with time. Renewal processes further generalise by allowing for ISI distributions other than the exponential. 

The quasi-periodic, anti-clustered nature of some volcanic seismicity indicates that in these cases, event occurrence is not independent, in the sense that the probability of an event falls after an event has just occurred. Similar properties of refractoriness have been observed in neuroscience \cite{barbieri} and cardiology \cite{heartbeats}. Successful modelling of such quasi-periodic data requires utilising appropriate ISI distributions.

Using Bayesian methods to fit point process models to the data allows us to incorporate prior knowledge with the information contained in the data, to obtain posterior distributions of the parameters. This enables us to examine the distributions of the parameters, and therefore to quantify the uncertainty of the estimates, and to examine the relationship between the parameters. These methods have also been shown to provide more reliable forecasts when data are sparse \cite{Boue2015}. We therefore use a Bayesian approach, although MLE methods could also be used \cite{chapter9}.

Here we describe and evaluate a variety of candidate point process models based on different ISI distributions to apply the FFM to a sequence of quasi-periodic LP earthquakes reported before a large explosive eruption at Tungurahua volcano. This paper aims to present statistical methodologies for examining the nature and role of the inter-event time distribution within the FFM framework, rather than evaluate the FFM itself. The FFM is already much discussed in the volcanic literature, but relatively few quantitative methods have been proposed to apply the model to data, and few studies have touched on the role of the inter-event time distribution and its significance for volcanic earthquake processes. Here we take a sequence of data where the suitability of the FFM (at least a power-law acceleration in mean earthquake rate with time) has already been established. We use these data to develop and test statistical methods, and test competing inter-event time distributions. The inter-event time distributions we choose are ones that have been proposed from related studies in different areas of statistical seismology, and we do not necessarily view them as an exhaustive list. We use Bayesian techniques, implemented through Markov chain Monte Carlo (MCMC). We find that although both gamma and inverse Gaussian distributions explain the data well, the gamma model provides more accurate estimates of the eruption time.

First we describe the LP earthquake data recorded before the July 2013 eruption at Tungurahua volcano, the model construction, and the methods used for parameter estimation and goodness-of-fit tests. We then apply the models to the data in both retrospective and forecasting scenarios, and compare goodness-of-fit and forecasting performance. Finally, we investigate reasons for the differing performance of different models, and implications for eruption forecasting based on quasi-periodic data.

\section{Data and model parameter estimation}

\subsection{Tungurahua volcano, monitoring data, and 14 July 2013 explosion}

Tungurahua is a large, active stratovolcano in the Eastern Cordillera of the northern Andes in Ecuador. After decades of unrest, a new eruption began in October 1999 \cite{samaniego}, characterized by vulcanian and strombolian episodes lasting a few months, and interspersed by months of quiescence \cite{Arellano2008,Bell2017}. On 14 July 2013, Tungurahua experienced a large vulcanian explosion, accompanied by the highest amplitude acoustic energy recorded at Tungurahua and a large gas plume. An ash column rose to 8.3km above the vent \cite{Hall2015}. The eruption was monitored by the seismic network of the Instituto Geofisco of the Escuela Polit\'{e}cnica Nacional (IGEPN). A sequence of small LP earthquakes were identified in the 24 hours before the explosion, best recorded at the short-period seismometer `RETU' located closest to the summit, and manually picked. When mean ISI approaches event duration, the chance of missing events begins to increase. Close to eruption, when mean ISI is significantly less than event duration, a large proportion of events are missed, and the sequence merges into tremor. Further work, perhaps involving picking events in simulated data, could quantify this effect more reliably.

Earthquake waveforms displayed a high degree of similarity, suggesting a closely co-located source, and implying that amplitude as recorded at RETU is a good proxy for relative energy. Amplitudes took a restricted range of values, suggesting a characteristic repeating source mechanism, and that the earthquake catalogue is largely complete \cite{bell2}. 

Average event rates, amplitudes, and energy release rates all increased towards the explosion, following the power-law form of the FFM \cite{bell2}. Earthquake inter-event times were quasi-periodic, and when re-scaled by average rate, closely approximated a Gamma distribution. 

\begin{figure}[!htbp]
\centering
\begin{minipage}[c][12cm][t]{.8\textwidth}
  \centering
  \caption{July 2013 data. LP earthquake rates preceeding 14th July 2013 eruption at Tungurahua volcano. Top panel: black line shows 15 minute event rates over time. Red line shows average ISI per 60 minute interval.Dashed lines show range of data used in fitting the model (second line is 200 minutes prior to eruption). Dotted line shows eruption time. Bottom panel:  spike plot, vertical lines show the times at which earthquakes were recorded\label{fig:data_plot}.}
  \centering
    \hspace*{-0ex}\includegraphics[width=0.82\textwidth]{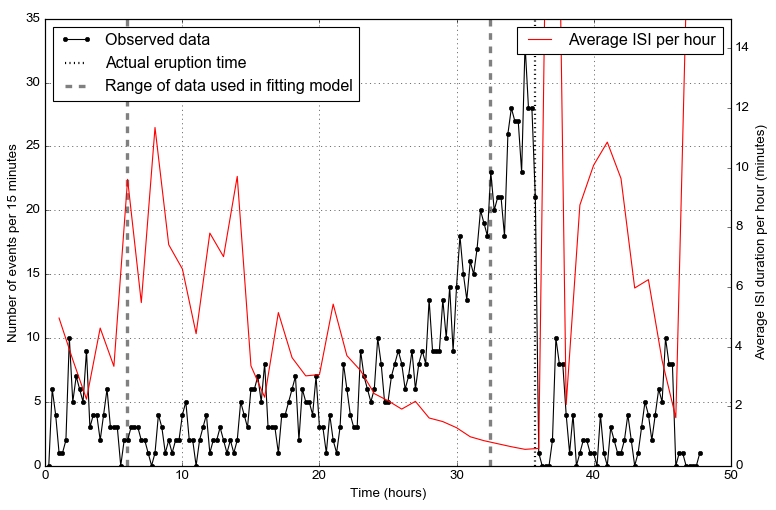}
  \centering
     \hspace*{0.5ex}\includegraphics[width=0.768\textwidth]{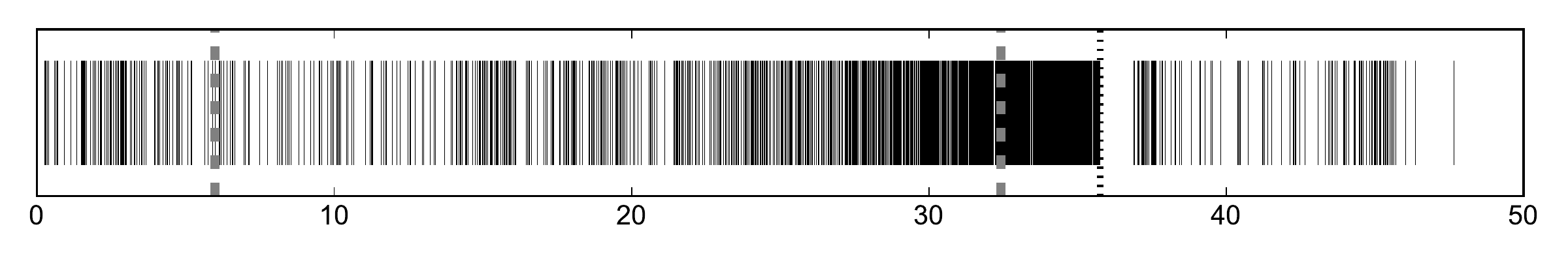}
     \end{minipage}
\end{figure}

The precursory LP earthquake sequence started at 06:00 on 13 July (all times are given in UTC) with the eruption starting effectively instantaneously at 11:46 on 14 July. Figure \ref{fig:data_plot} shows the 15 minute event rate, and a plot of the `spikes' (vertical lines demonstrate times at which LP earthquakes were observed). The event rate grows at an increasing rate up to the eruption, with the ISI decreasing from over 10 minutes to below 30 seconds, and a total of just over 1,000 events. Within 200 minutes of the eruption time, the rate of earthquakes becomes sufficiently high that consecutive events being to merge in to one another. This high rate means that it is likely that some earthquakes are missed from the catalogue. The ISIs demonstrate quasi-periodic behaviour, being more regular than would be seen if the events followed a Poisson process, and thus not independent \cite{bell2}. As such, this sequence provides an ideal dataset to develop a methodology for comparing different inhomogeneous point process models.

\subsection{Parameter estimation using MCMC}

We use a Bayesian approach to fit model parameters to the data, i.e. specifying prior distributions on the parameters which contain known information, and calculating the likelihood of the data under each model. This allows us to incorporate prior knowledge for inference when it is available (or as it becomes available). Then the posterior distributions of the parameters given the data are obtained. As these calculations are very complex (or impossible) to do exactly, a computational approach is used instead. Markov Chain Monte Carlo techniques use random numbers to sample repeatedly from the posterior distribution and thus obtain summary statistics of interest without having to calculate the posteriors directly.

Model parameter estimation was performed using MCMC implemented through the PyMC3 package \cite{Salvatier2016}. The No-U-Turn Sampler (NUTS) \cite{nuts} was used, initialised with the automatic differentiation variational inference algorithm (ADVI) \cite{ADVI}. This was found to result in quick convergence, as assessed using standard tests, with the chains demonstrating good mixing. PyMC3's built in tools were used to check performance. Initialising the chain at 10 randomly selected initial points and calculating the Gelman-Rubin statistic \cite{gelmanrubin}, and running the chain for a large number of iterations, showed no apparent issues with convergence. The chain was run for 20,000 iterations, discarding a burn-in of 1,000. 

Prior beliefs about the parameters of the power law event rate in (\ref{lambda}), such as the likely range of values and the mean, are incorporated into prior distributions. For constructing these, there is no general rule which would help narrow the range for the more likely values of $k$ or $t_f$, so bounded uniform priors were used. However, there are prior beliefs regarding the power law coefficient $p$ which can be used. In the literature, the likely range of $p$ is generally given as 0.5 to 1.9 \cite{ogata1999,utsu1995,wiemer1999}, with a value near 1 being most likely. This information is reflected in the lognormal prior used:
\[ p \sim \text{logN}(0.1,0.25) \]

The likelihood is calculated using the observed data and the constructed ISI pdf. Then, MCMC methods are applied to produce posterior estimates of the parameters. 

\section{Model construction}

We construct a number of models that describe the ISI distributions of the event times, with a Markov dependence on spiking history. The choice of these distributions was guided by considering which shapes of the ISI pdf would be reasonable for the data at hand, as well as considering available literature on analysing the fit of several distributions to model (non-volcanic) earthquake event times. Four models are chosen for the comparison: the inhomogeneous Gamma (IG), inhomogeneous Poisson (IP), inhomogeneous Weibull (IW), and inhomogeneous inverse Gaussian (IIG).

The IG model was proposed by Bell et al. (2018). It generalises the IP model which is commonly used to model independent earthquake events, by having an extra shape parameter $\alpha$ (the IP is obtained from the IG model by setting $\alpha = 1$). $\alpha$ relates to periodicity of the data, which is defined as the ratio of the mean and standard deviation, which is $\sqrt{\alpha}$ for the IG model. Thus, it allows us to check whether introducing the $\alpha \neq 1$ parameter results in a fit that is significantly better, and thus to see whether the data is quasi-periodic.

The IW model is chosen as it also simplifies to the IP model by setting the shape parameter to 1. The IIG model is chosen as the shape of the resulting ISI probability density appears similar to that of the IG model. This distribution has also been used previously to study (non-volcanic) earthquake occurrence \cite{matthews2002}.

Details of the construction of the models are given in the following sections, following the methodology of \cite{barbieri,timerescaling,chapter9}.

\subsection{Inhomogeneous Gamma}

First, we construct an inhomogeneous renewal process with gamma ISIs. The gamma probability density with shape parameter $\alpha>0$ is given by
\[
h(z) = \frac{1}{\Gamma(\alpha)} z^{\alpha-1} \exp(-z)
\]
where $\Gamma(\alpha)$ is the gamma function, $z>0$. Using (\ref{lambda}), the intensity rescaling transformation is \cite{barbieri,timerescaling}
\begin{equation} \label{intensity_rescaling}
z_m = g(s | s_m) = \alpha \int^{s}_{s_{m}} \lambda(u) \dd{u},\; s > s_m
\end{equation}
where $s_m$ is the time of the $m$-th event. 
Using the change of variables formula \cite{port}
\[
f(s | s_m) = \left| \frac{\dd{g}}{\dd{s}} \right| h(g(s | s_m))
\]
results in the inhomogeneous gamma (IG) probability density for spike times of \cite{barbieri}:
\begin{equation}\label{rate}
f(s | s_m) = \frac{\alpha \lambda(s)}{\Gamma(\alpha)} \left[ \alpha \int^{s}_{s_{m}} \lambda (u) \dd{u} \right]^{\alpha -1} \exp \left( -\alpha \int^{s}_{s_{m}} \lambda (u) \dd{u} \right), \; s > s_m
\end{equation}

\subsection{Inhomogeneous Poisson}

The IP process is a special case of the IG process, with $\alpha = 1$, giving the spike time probability density \cite{barbieri}
\[
f(s| s_{m}) = \lambda(s) \exp \left( -\int^{s}_{s_{m}} \lambda(u) \dd{u} \right), \; s > s_m
\]

\subsection{Inhomogeneous Weibull}

The Weibull density with zero scale and location parameters and shape parameter $\phi$ is given by
\[
h(z) = \phi z^{\phi - 1} \exp(-z^\phi), \; z>0, \phi>0
\]

With the transformation
\[
z_m = g(s | s_m) = \phi \int_{s_m}^s \lambda(u) \dd{u}, \; s>s_m
\]
the spike time probability density is:
\[
f(s | s_m) = \phi^2 \lambda(s) \left( \phi  \int^{s}_{s_{m}} \lambda(u) \dd{u} \right)^{\phi-1} \exp \left[ - \left( \phi  \int^{s}_{s_{m}} \lambda(u) \dd{u} \right)^{\phi} \right].
\]

\subsection{Inhomogeneous inverse Gaussian}

From Barbieri et al. (2001),
\[
h(z) = \left( \frac{1}{2 \pi z^3} \right)^{\frac{1}{2}} \exp \left[ - \frac{1}{2} \frac{(z-\psi)^2}{\psi^2 z} \right], \; z>0, \psi>0
\]
With the transformation (different to the IG and IW models)
\[
g(s | s_m) = \int_{s_m}^s \lambda(u) \dd{u}
\]
the IIG spike time probability density is given by
\[
f(s | s_{m}) = \frac{\lambda(s)}{\left[ 2 \pi \left( \int^{s}_{s_{m}} \lambda(u) \dd{u} \right)^3  \right]^{\frac{1}{2}}} \cdot \exp \left( -\frac{1}{2} \frac{\left( \int^{s}_{s_{m}} \lambda(u) \dd{u} - \psi \right)^2}{ \psi^2  \int^{s}_{s_{m}} \lambda(u) \dd{u}} \right).
\]

\section{Goodness-of-fit tests}

The goodness-of-fit of the models is assessed to determine how well the data is described by the candidate models. Q-Q and K-S plots are used, as these allow us to explore which characteristics of the data the models describe well \cite{barbieri}.

Both of these techniques are constructed using the rate rescaling theorem \cite{ogata1988,barbieri,timerescaling}. The conditional intensity is defined as:
\[
r(s) = \frac{f(s | s_{m-1})}{1 - \int^s_{s_{m-1}} f(u|s_{m-1})\dd{u}}
\]
The rate rescaling theorem then states that the transformed quantities
\[
\tau_m = R(s_m) - R(s_{m-1}), \; \text{where } R(s) =  \int^{s}_{0} r(u) \dd{u}
\]
are independent and identically distributed (i.i.d.) exponential with rate 1. 

In testing goodness-of-fit, the posterior mean parameter values were used; all four of the posterior distributions were found to be unimodal and quite symmetric, therefore very similar results would be obtained if using other point estimates such as the mode.

\subsection{Q-Q plots} \label{QQsection}

The Quantile-Quantile (Q-Q) plot is a method of assessing the goodness of fit of a model \cite{wilks}, by using the rate rescaling theorem and comparing the values of the expected model quantiles to the empirical quantiles. This is constructed using the following method \cite{barbieri,chapter2}. Let $M$ be the total number of observed spikes. As noted above, if the behaviour is appropriately described by the model, then we would expect the $\tau_m$ to follow an exponential distribution with rate 1, with cdf $F_\tau (\tau) = 1-e^{-\tau}$, and pdf $f_\tau (\tau) = e^{-\tau}$. 

To construct the Q-Q plot, the $\tau_m$ are listed in increasing order, as $\tau_{(m)}$. Let $b_m = (m-\frac{1}{2})/M$ for all $m =1 \hdots M$. Then the $\tau_{(m)}$s are the empirical quantiles, which are compared to the model quantiles, calculated as:
\[
\tilde{\tau}_m = F^{-1}_{\tau} (b_m) = -\log(1-b_m)
\]
If the model fits well, we would expect the points to adhere to a diagonal line with slope 1 \cite{wilks}.

\subsection{K-S plots}

This Kolmogorov-Smirnov (K-S) goodness of fit method \cite{barbieri,ogata1988,chapter2} is applied next to analyse the fit of the model. If the model fits the data appropriately, then the $\tau_m$ are i.i.d. exponential with rate 1. Then $u_m = 1-e^{-\tau_m}$ are i.i.d. Uniform(0,1). Therefore, sorting in increasing order to produce $u_{(m)}$, we would expect a plot of $u_{(m)}$ against $b_m$ to be a diagonal line. 

Error bounds for the K-S plot are constructed based on the critical values of the Kolmogorov-Smirnov test statistic \cite{zwillinger}.

\subsection{Comparison with simulated data}

The error bounds shown on the K-S plots provide an easy way of gauging the significance of the deviations for each model and comparing their goodness-of-fit. For the Q-Q plots, to avoid computing pointwise error bounds for all models individually, we can estimate the degree of expected deviation from the diagonal by simulating datasets. As the simulated points are generated from the exact distributions, we would expect them to adhere very well to the diagonal line, and therefore we can gauge the degree of natural variation that can be expected by data coming from the point process. 

Data is simulated using the intensity rescaling transformation, by solving (\ref{intensity_rescaling}) for $s_m$:
\begin{equation} \label{tk}
s_m = t_f - \left[ \frac{z(p-1)}{\alpha k} + (t_f-s_{m-1})^{1-p} \right]^{\frac{1}{1-p}}
\end{equation}
and thus constructing the following algorithm:
\begin{enumerate}
\item Initialise $m=1$, $s_{0} = 0$, $T = \{ \emptyset \}$. Set $s_{\text{end}}$ to the end time for the simulated data. 
\item Randomly draw $z \sim \Gamma(\alpha,1)$. 
\item Solve (\ref{tk}) to obtain $s_m$. 
\item If $s_m<s_{\text{end}}$: add $s_m$ to $T$, set $s_{m-1} = s_m$ and return to step 2. Otherwise, stop and output $T$.
\end{enumerate}

Data can be simulated similarly for the IG and IIG models, but only simulations from the IG model will be used further in this paper.

\section{Application to real data -- retrospective \label{retro}}

For fitting the model, spike times were used from the start of quasi-periodic spiking activity to a threshold $t_{\text{end}}$ of 200 minutes before the eruption (where event merger begins to occur); these bounds are shown as grey dashed lines in Figure \ref{fig:data_plot}.  

The posterior statistics for the parameters for each model are given in Table \ref{stats1} below. In each case, the resulting distributions are unimodal, with reasonably narrow highest posterior density intervals (HPDIs), i.e. the narrowest 95\% credible intervals. It was found that $k$, $t_f$ and $p$ appear to be strongly correlated with each other. These parameters do not appear to be significantly correlated with $\alpha$, $\phi$ and $\psi$, for the IG, IW and IIG models, respectively. 

\begin{table}[h!]
\centering
\caption{Posterior statistics for parameters of the IG, IP, IW and IIG models. \label{stats1}}
\begin{tabular}{| l | l | r | r | c |} \hline
\bf{Model } & \bf{Parameter}       & \bf{Mean}    & \bf{Std. dev.} & \bf{95\% HPDI}              \\ \hline
IG & $k$      & 295.382 & 29.890  & {[}247.645, 357.768{]} \\
& $t_f$    & 1.321   & 0.070   & {[}1.197, 1.464{]}     \\
& $p$      & 1.286   & 0.189  & {[}0.950, 1.680{]}    \\ 
& $\alpha$ & 2.256   & 0.120   & {[}2.025, 2.492{]}     \\ \hline

IP & $k$      & 302.274 & 42.403  & {[}234.060, 391.955{]} \\
& $t_f$    & 1.326   & 0.089   & {[}1.177, 1.506{]}     \\
& $p$      & 1.289   & 0.237   & {[}0.876, 1.771{]}    \\ \hline

IW & $k$      & 171.303 & 21.629  & {[}139.012, 212.668{]} \\
& $t_f$    & 1.331   & 0.071   & {[}1.210, 1.474{]}     \\
& $p$      & 1.396  & 0.202   & {[}1.025, 1.799{]}    \\ 
& $\phi$ & 1.487 & 0.044 & [1.402, 1.572] \\ \hline

IIG & $k$      & 231.646 & 30.762  & {[}184.073, 287.225{]} \\
& $t_f$    & 1.301   & 0.093   & {[}1.153, 1.484{]}     \\
& $p$      & 0.970  & 0.190   & {[}0.633, 1.342{]}    \\ 
& $\psi$ & 0.665 & 0.044 & [0.584, 0.754] \\ \hline
\end{tabular}
\end{table}

\subsection{Q-Q plot analysis}

\begin{figure}[!h]
  \caption{Q-Q plot for IP, IG, IIG and IW models. A square root transformation has been applied to correct for the mass of the points being very concentrated near zero. \label{fig:QQcomparison} }
  \centering
    \includegraphics[width=0.6\textwidth]{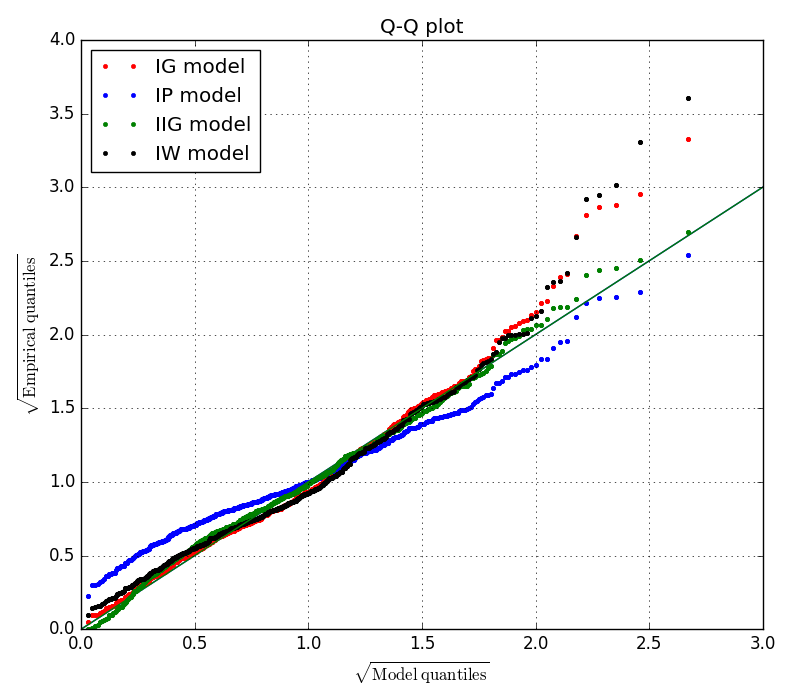}
\end{figure}

Figure \ref{fig:QQcomparison} shows the Q-Q plots of all four models, constructed as described in Section \ref{QQsection}. As the mass of the points is very concentrated near zero (with the distribution being highly skewed), a square root transform has been applied to $\tilde{\tau}_m$ and $\tau_{(m)}$ to mitigate this and make the plots easier to interpret.

The IG model (red points) appears to fit well, with most of the points adhering closely to the diagonal line. There are several outliers towards the tail of the plot (representing approximately 5\% of the data), where the points deviate significantly from the diagonal, suggesting that the model underestimates the probability of the data in the higher percentiles. These mostly correspond to spikes which have particularly long ISIs preceding them (falling in the right tails of their corresponding ISI densities).  This could be due to a number of reasons:
\begin{itemize}
\item The model could inadequately describe the data, missing out key characteristics that affect ISI durations.
\item Spikes missing from the event catalogue during picking (e.g. due to them having lower amplitudes and being difficult to distinguish from the background signal). A plausible picking error rate of around 5\% would result in several anomalously long ISIs, which in reality contained a spike inbetween. 
\item There could be other effects of the physical system which would result in the sporadic occurrence of longer ISIs than usual (i.e. to do with the underlying physical processes that differ from the simplified 'ideal' trend). 
\end{itemize}

The IW model (in black) appears to fit most of the points as well as the IG model, however its outliers are more extreme, with poorer adherence to the diagonal in the lower and higher percentiles. 

The IP model (in blue) does not fit the data well, showing significant deviation from the diagonal line for most of the spikes, both in the lower percentiles and towards the tail. 

The IIG model, however, appears to fit the data better than the IG model, particularly in high model quantiles. The green points adhere closer to the diagonal than the red points, and while there are still some outliers, these lie closer to the diagonal. This suggests that to some extent, the IIG model describes the observed data better.

\subsection{K-S plot analysis}

\begin{figure}[!h]
  \caption{K-S plot for IP, IG, IIG and IW models, with 95\% and 99\% error bounds. \label{fig:KScomparison}}
  \centering
    \includegraphics[width=0.6\textwidth]{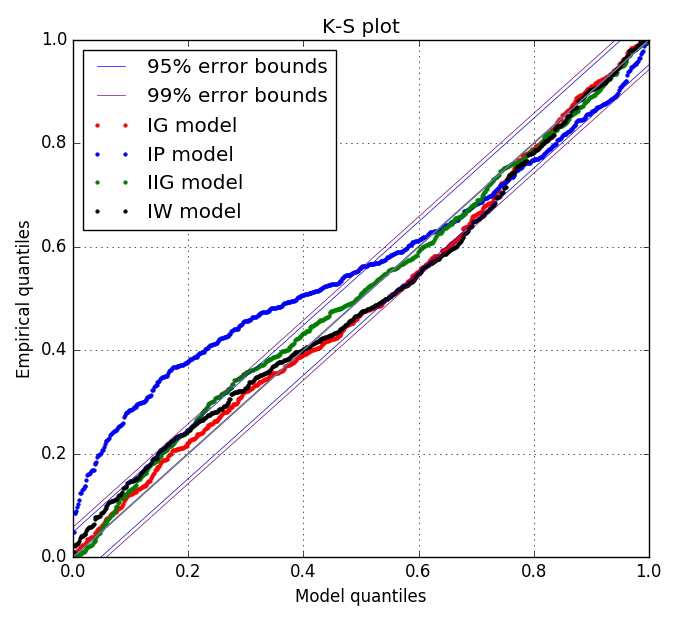}
\end{figure}

In the K-S plot in Figure \ref{fig:KScomparison}, the IP model clearly does not lie within the 95\% or 99\% bounds for a significant proportion of the points. The IG model appears to fit well, mostly adhering to the diagonal very well, staying within the two-sided 95\% error bounds, apart from some slight ``bowing" in the middle which is where the only failure occurs. This implies that the model underestimates the probability of some of the data in the middle quantiles. The IIG model, however, appears to fit within the bounds better than the IG model. The IW model demonstrates a similar but slightly poorer fit to the IG model. This confirms the findings from the Q-Q plot analysis. 

\subsection{Independence of rescaled ISIs}

The Q-Q and K-S plots presented above allow for a visual check of whether the proposed models produce significant deviations from the assumption that the time-rescaled process is Poisson with rate 1, by checking whether $\tau_m$ and $u_m$ appear to follow the expected exponential and uniform distributions, respectively. In this section we look at other properties of the rescaled ISIs to further check for inconsistencies with the assumption that the time-rescaled process is Poisson with rate 1.
\begin{figure}[!htbp]
  \centering
  \caption{}
\begin{subfigure}[!htbp]{0.55\linewidth}
  \subcaption{Plots showing $u_m$ against $u_{m+1}$, i.e.``neighbouring" rescaled time intervals. If a clear pattern is noted, this would suggest lack of independence between consecutive transformed ISIs. \label{fig:neighbours}}
  \centering
    \includegraphics[width=\textwidth]{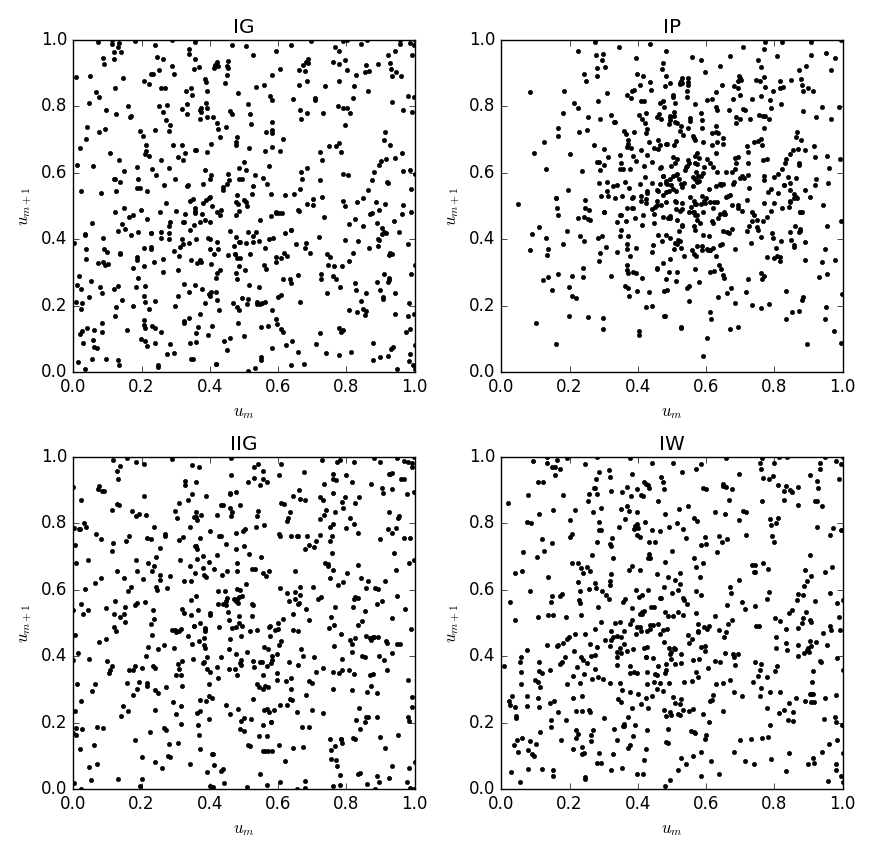}
\end{subfigure}
\begin{subfigure}[!htbp]{0.9\linewidth}
  \subcaption{Autocorrelation plots for each of the models. Autocorrelation is calculated for (time-ordered) $\tau_m$ and shown for lags of up to 100. The solid grey and solid red horizontal lines show the 95\% and 99\% confidence bands, respectively. \label{fig:autocorr_plot}}
  \centering
    \includegraphics[width=\textwidth]{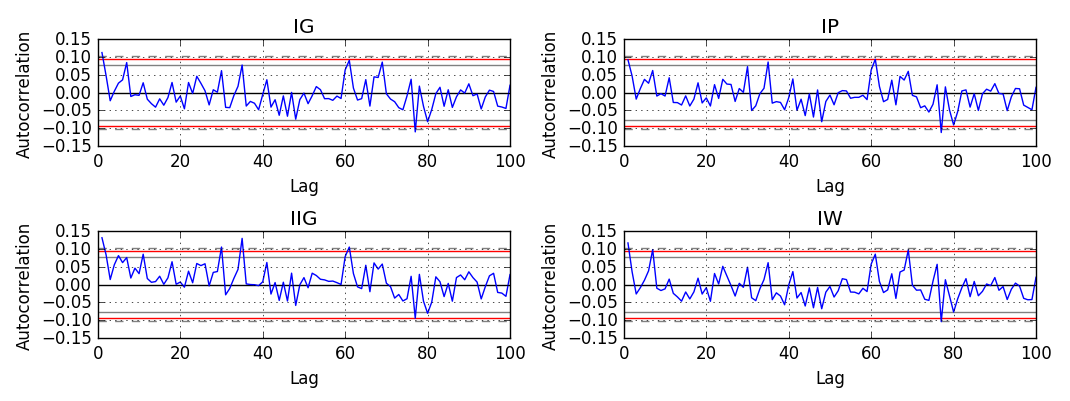}
\end{subfigure}
\end{figure}

Some examples of such checks are provided by Ogata (1988), which include testing independence of the transformed ISIs: Ogata plots neighbouring intervals against each other (i.e. plotting $u_m$ against $u_{m+1}$) to check for trends -- strong patterns would suggest that there is a dependence between consecutive ISIs that is not appropriately accounted for by the models. We follow this suggestion, in Figure \ref{fig:neighbours}. Moreover, we also show an autocorrelation plot for each of the models in Figure \ref{fig:autocorr_plot}, to check if there is a dependence between intervals at higher lags.

The additional plots do not appear to show any serious deviation from the assumptions of independence for the rescaled quantities for the IG, IIG or IW models. For the IP model, there appears to be some clustering in Figure \ref{fig:neighbours}, and very sparse regions at low values of $u_m$ -- again suggesting lack of fit for this model.

\subsection{Further analysis of fit of IG model}
The IG model is proposed and fitted to the same data by Bell et al. (2018), and will be discussed in further sections in detail in the context of forecasting the eruption. As previously noted, the IG model generalises the IP model by introducing the $\alpha$ parameter which relates to periodicity and allows us to check whether the data is quasi-periodic. The IG model is therefore of particular interest, and in this section we further consider the fit of this model.

\begin{figure}[htbp]
  \caption{Histogram of transformed quantities $z_k = \alpha \int^{t_k}_{t_{k-1}} \lambda(t) \dd{t}$, showing the fitted gamma distribution (using posterior mean parameter values). \label{fig:z_hist}}
  \centering
    \includegraphics[width=0.6\textwidth]{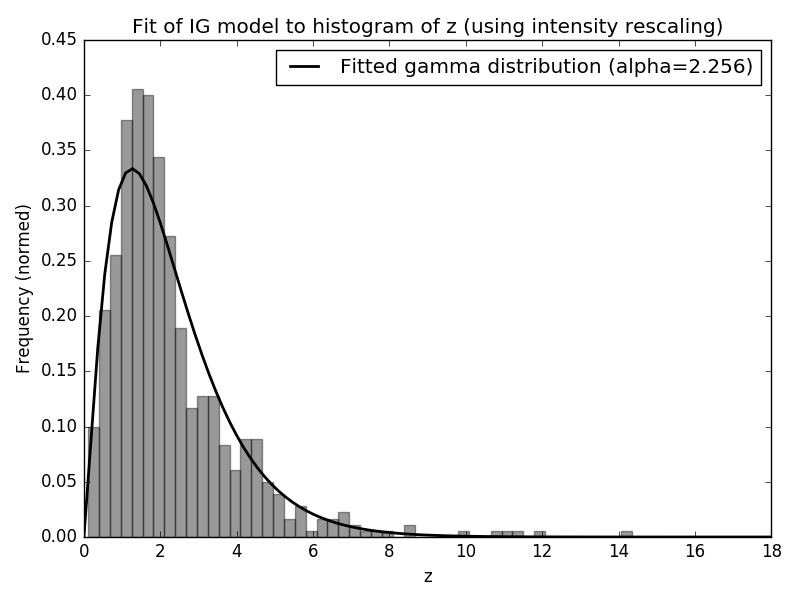}
\end{figure}

Figure \ref{fig:z_hist} shows the fit of the IG model using the intensity rescaling transformation as given in (\ref{intensity_rescaling}), giving a histogram of $z_k = \alpha \int^{t_k}_{t_{k-1}} \lambda(t) \dd{t}$, and showing the fitted gamma distribution (using the posterior mean of $\alpha$). 

There are more values in the right tail of the distribution than accounted for by the IG model. This implies that the model underestimates the probability of some long ISIs, consistent with the findings of the Q-Q plot analysis. 

Moreover, there is a lack of fit around the mode, with more values observed than would be expected under the gamma distribution. This is consistent with the ``bowing" effect seen in the middle quantiles of the K-S plot, as having a larger than expected number of observations in any area of the histogram would make the line of the K-S plot dip further below the diagonal. This is counterbalanced by slightly fewer observations than expected to the right of the mode, which would cause the line in the K-S plot to lift back up. 

\subsection{Goodness of fit using simulations} \label{gof_sim}

\begin{figure}[!h]
  \caption{Q-Q (left panel) and K-S (right panel) plots for 10 simulated datasets. Simulated datapoints are shown in orange; points for IG (red), IP (blue), IIG(green) and IW (black) models are also shown.\label{fig:simQQKScomp}}
  \centering
    \includegraphics[width=0.85\textwidth]{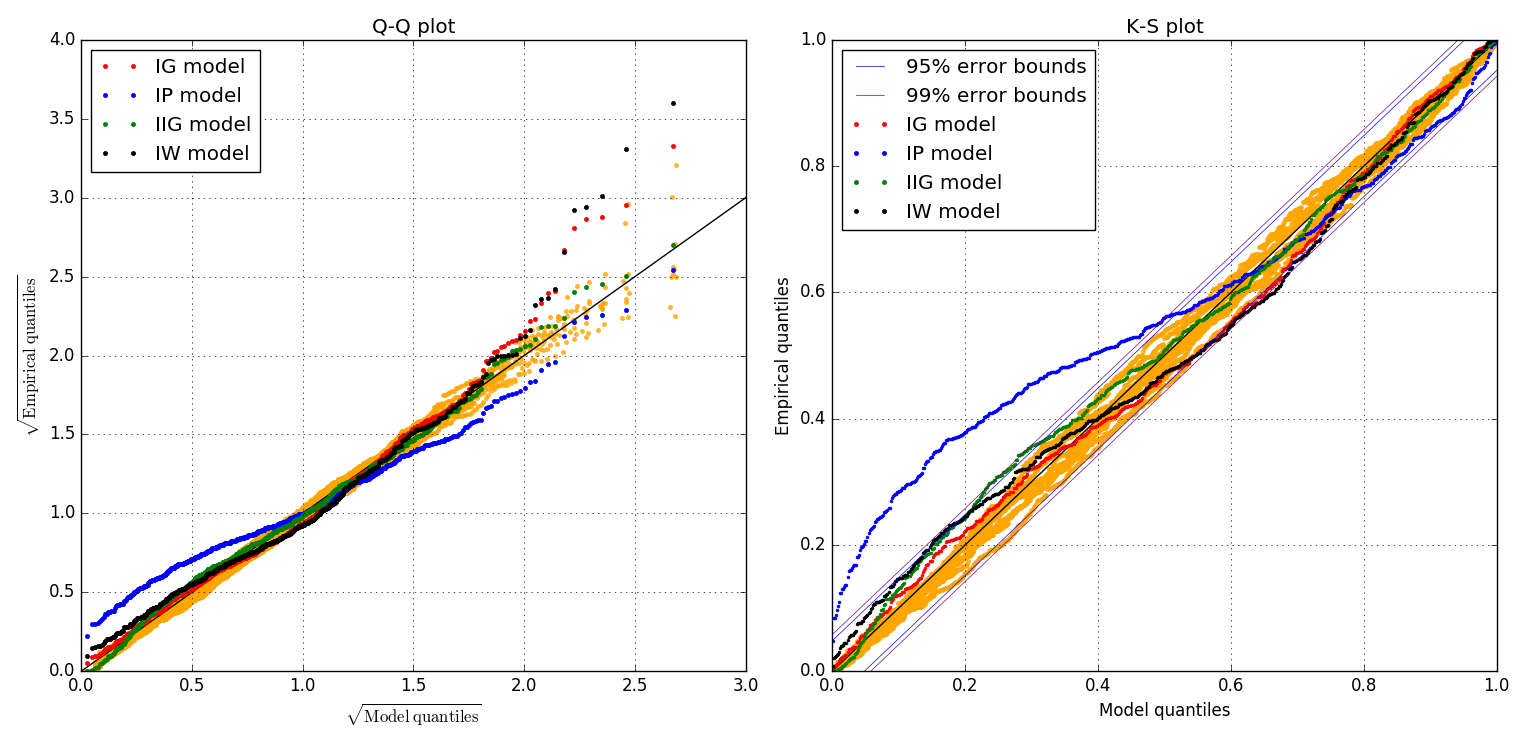}
\end{figure}

Ten datasets were simulated from the IG model, using parameter values given by random samples from the MCMC chain. The Q-Q and K-S plots were constructed using the method previously described, with the simulated data shown as orange points in Figure \ref{fig:simQQKScomp}. The IG (red), IW (black), IIG (green) and IP (blue) points are also shown. 

The simulated data displays some variation from the line in the Q-Q plot, particularly in the higher percentiles. The IIG points appear to be very close to the simulated data, displaying outliers that are very similar to those from the simulated datasets. The IG and IW points appear to fit less well, clearly having outliers more extreme than any of the simulated datasets. The K-S plot demonstrates that the simulated data lies almost entirely within the error bounds, as expected.

\begin{figure}[!h]
  \caption{Q-Q (left panel) and K-S (right panel) plots for 10 simulated datasets, with 5\% of the simulated data removed uniformly over time. Simulated datapoints are shown in orange; points for IG (red), IP (blue), IIG(green) and IW (black) models are also shown.\label{fig:simQQKScomp_missing}}
  \centering
    \includegraphics[width=0.85\textwidth]{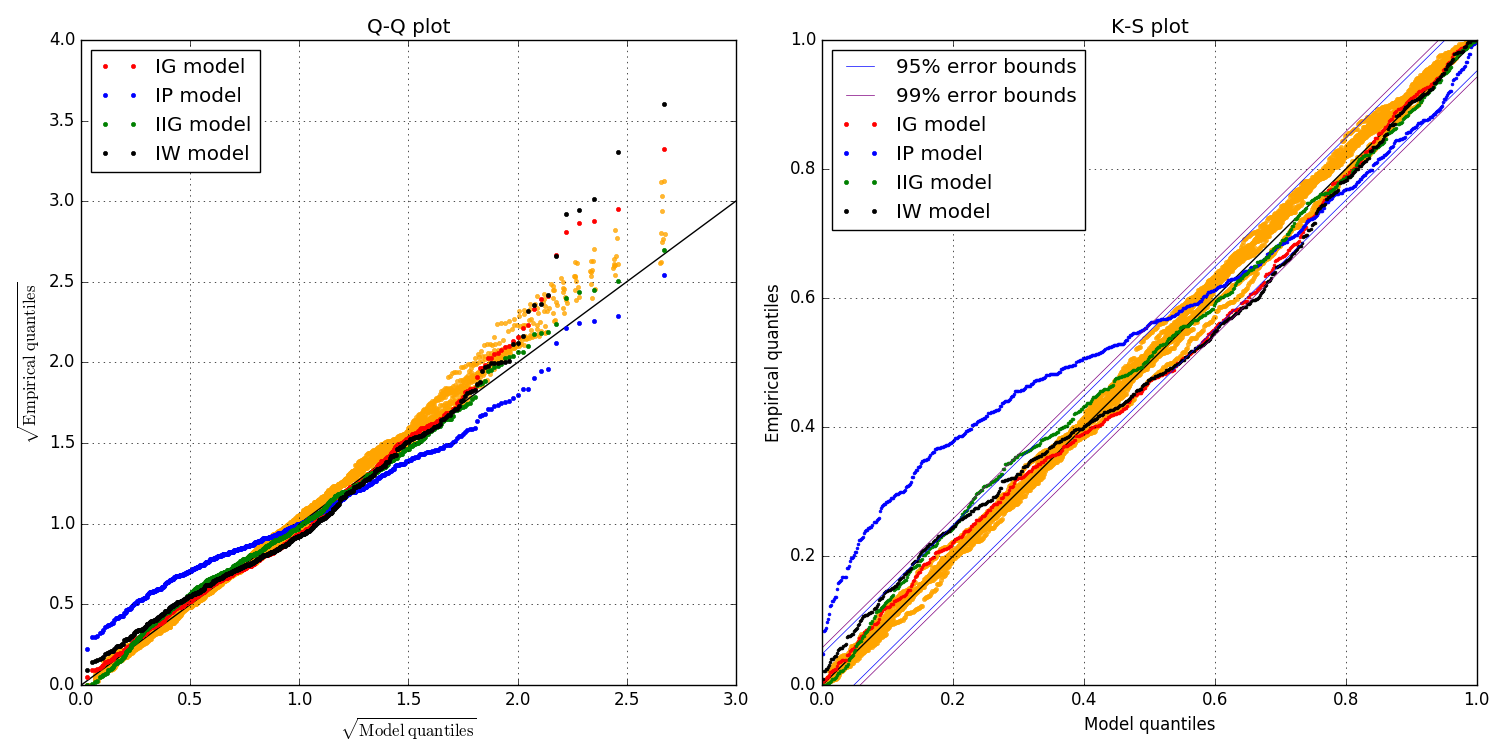}
\end{figure}

Incomplete synthetic catalogues provide a further test as to whether the outliers towards the tail of the Q-Q plots are likely to be due to some events being missed out when picking the catalogue from the primary seismic data. 10 datasets were simulated and then 5\% of the spikes were removed uniformly over time (being the approximate expected rate of errors). The Q-Q and K-S plots were then again constructed, as shown in Figure \ref{fig:simQQKScomp_missing}. 

The resulting plot demonstrates that the incomplete datasets adhere less well to the diagonal line in the Q-Q plot, but in particular the points towards the tail are significantly higher above the line than for the unaltered simulated data. This brings the simulated points closer towards the outliers of the IG and IIG models. Indeed, the IG and IIG models' outliers now appear to be very similar to the simulated data. This suggests that some of the data being missed out during collection could account for the outliers of the IG and IIG models.

\section{Application to real data -- forecasting}

We evaluate the potential utility of the candidate FFM models for providing accurate forecasts of eruption time and other parameters. A pseudo-prospective forecast method was used to review the evolution of posterior distribution of $t_f$ over time, up to the eruption \cite{bell2}. We use the term `pseudo-prospective' to refer to simulated forecasting, where model parameters including the eruption time are variables in our inversions. The observed time series was split into 50 intervals of equal duration, and the MCMC chain rerun for each interval, incrementally adding data from the next time interval at each iteration.

\subsection{Forecast of $t_f$}
The estimate of the eruption time is given by the parameter $t_f$. As the actual eruption time for the observed dataset is known retrospectively (as 1.241 days from the start of activity), we can evaluate how soon after the start of LP earthquake activity each model starts providing reasonable estimates of the upcoming eruption time, and how precise they are. 

\begin{figure}[!htbp]
  \centering
  \caption{Evolution of $t_f$ forecast for observed data, showing mean and 95\% HPDIs for $t_f$, estimated using increasing subsets of the data up to eruption. Horizontal green line represents the actual eruption time. Vertical dashed pink line shows threshold used in fitting models in Section \ref{retro}, of 200 minutes before the eruption (data before this line was used).}
\begin{subfigure}[!htbp]{0.9\linewidth}
  \subcaption{IG, IP and IW models.\label{fig:fc_observed_IG_IP_IW}}
  \centering
    \includegraphics[width=\textwidth]{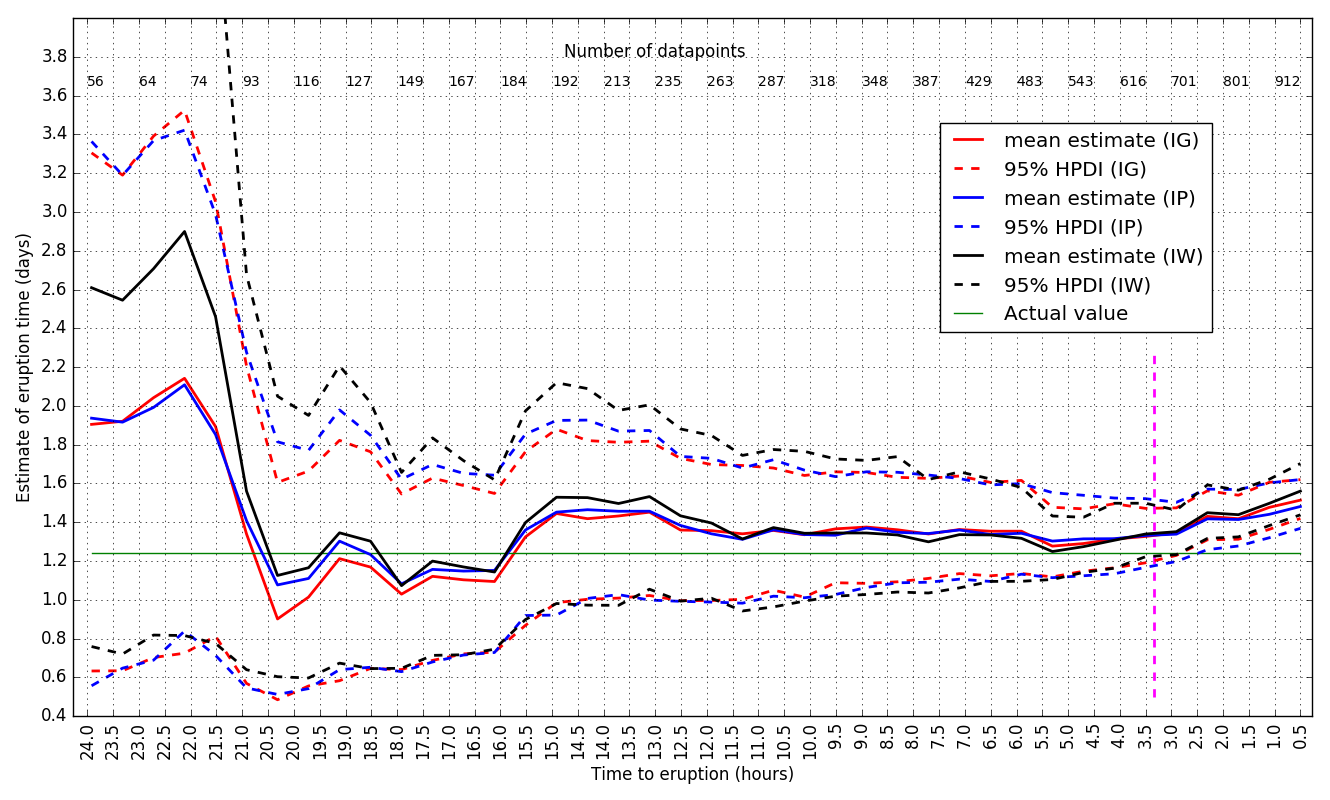}
\end{subfigure}
\\ \vspace{10pt}
\begin{subfigure}[!htbp]{0.9\linewidth}
  \subcaption{IIG model.\label{fig:fc_observed_IIG}}
  \centering
    \includegraphics[width=\textwidth]{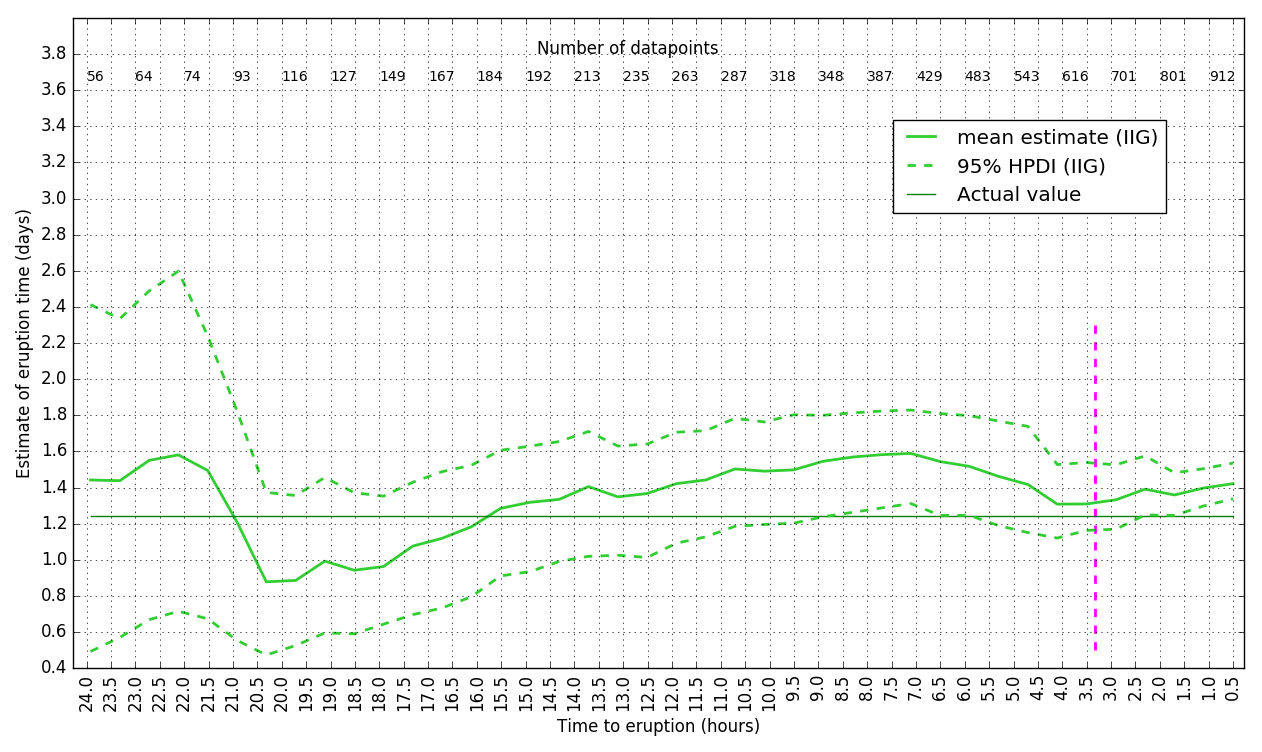}
\end{subfigure}
\end{figure}

Figure \ref{fig:fc_observed_IG_IP_IW} shows the resulting evolution of the $t_f$ approximation for the IG and IP models, and Figure \ref{fig:fc_observed_IIG} shows that for the IIG model. For all models, the forecasts start out with very wide credible intervals, when few data points are available. Around 15 hours before the eruption, when around 190 events have been recorded, the mean of the forecast time reaches a value of around 1.5 and stops fluctuating as strongly over the next few hours, converging to a value close to the true eruption time. Within around 4 hours before the eruption, the estimate again starts to diverge away from the actual eruption time, overestimating it.

Although the IP model was found to provide a much worse retrospective fit to the data, it provides forecasts for $t_f$ which are very close to those of the IG model. This is explained by the fact that $k, t_f$ and $p$ are mostly uncorrelated with $\alpha$, therefore their forecasts are not affected significantly by setting $\alpha = 1$. Thus, for this dataset, for the purpose of predicting the eruption time while events are being recorded, the IP model is appropriate. As the IP model has fewer parameters than the other models, calculations are faster. 

The IW model provides similar mean forecasts for $t_f$ throughout, however with a significantly wider 95\% HPDI than the IG or IP models up to around 8 hours before the eruption. Therefore, even though the model has one more parameter than the IP model, it is worse at forecasting the eruption time. 

Although the IIG model was found to fit the observed data better, the forecasts of $t_f$ are consistently worse than for the other models, until very close to the true eruption time. The model overestimates $t_f$ consistently, and gives wider credible intervals that the other models. Therefore the IIG model is not useful for forecasting purposes. This may be due to the shape of the resulting ISI distributions, with the parameter estimates becoming reasonably accurate only when the event rate observed in the data increases significantly, indicating that an eruption is near.

\subsection{Estimate of periodicity}

Periodicity is defined as the ratio of the mean and standard deviation. For earthquakes clustered in time, periodicity will be less than 1 as the variance of ISIs will be relatively high; for highly periodic earthquakes, the variance of ISIs will be low, resulting in periodicity greater than 1 \cite{Bell2017}. All of the models (IP, IG, IW and IIG) assume a constant periodicity. 

For the IG model, the periodicity is equal to $\sqrt{\alpha}$. The evolution of periodicity with time for the IG model forecast shows an approximately linear increase from around 13 hours before eruption \cite{bell2}. The models all assume that periodicity is constant throughout, so we would expect the forecast of $\sqrt{\alpha}$ to remain relatively stable. However, given the low correlation between periodicity and the other parameters, this finding is not expected to affect the values of the other parameters significantly.

The same is observed for the IIG model, which has periodicity $\frac{1}{\sqrt{\psi}}$, 
increasing linearly from around 13 hours before eruption. As with the IG model, however, the correlation between periodicity and the other parameters is low.

\subsection{Merging events and catalogue incompleteness} \label{merging}

The pink dashed line in Figure \ref{fig:fc_observed_IG_IP_IW} shows the time threshold at which the retrospective analyses were undertaken (200 minutes before the true eruption time). Up to this point, the forecast gradually gets closer to the actual value of $t_f$, with the nearest estimate made approximately 5.3 hours before the eruption for the IG and IP models. However, as more data is added nearer the eruption, the estimates diverge from the true value, with the mean estimate of $t_f$ increasing to around 1.5, half an hour before the eruption. The same effect is observed from looking at the evolution of the estimate of the power-law parameter $p$ within each of the models, using increasing subsets of data; the mean forecast remains relatively stable at around 1.2, up until around 3 hours before the eruption, when it starts climbing steadily.

\begin{figure}[!h]
  \caption{Event rate plots showing effect of merging event times. Left panel shows number of events per 15 minutes, for observed data, and for IG and IIG models fit using data up to 100 minutes and 200 minutes before the eruption. Right panel shows the same event rates on a log scale. Orange vertical line shows the actual eruption time. Dotted lines show the threshold of 100 and 200 minutes before the eruption. Dashed light and dark blue lines show the eruption times as predicted by the IG and IIG model, respectively.\label{fig:IG_IIG_incompleteness}}
  \centering
    \includegraphics[width=\textwidth]{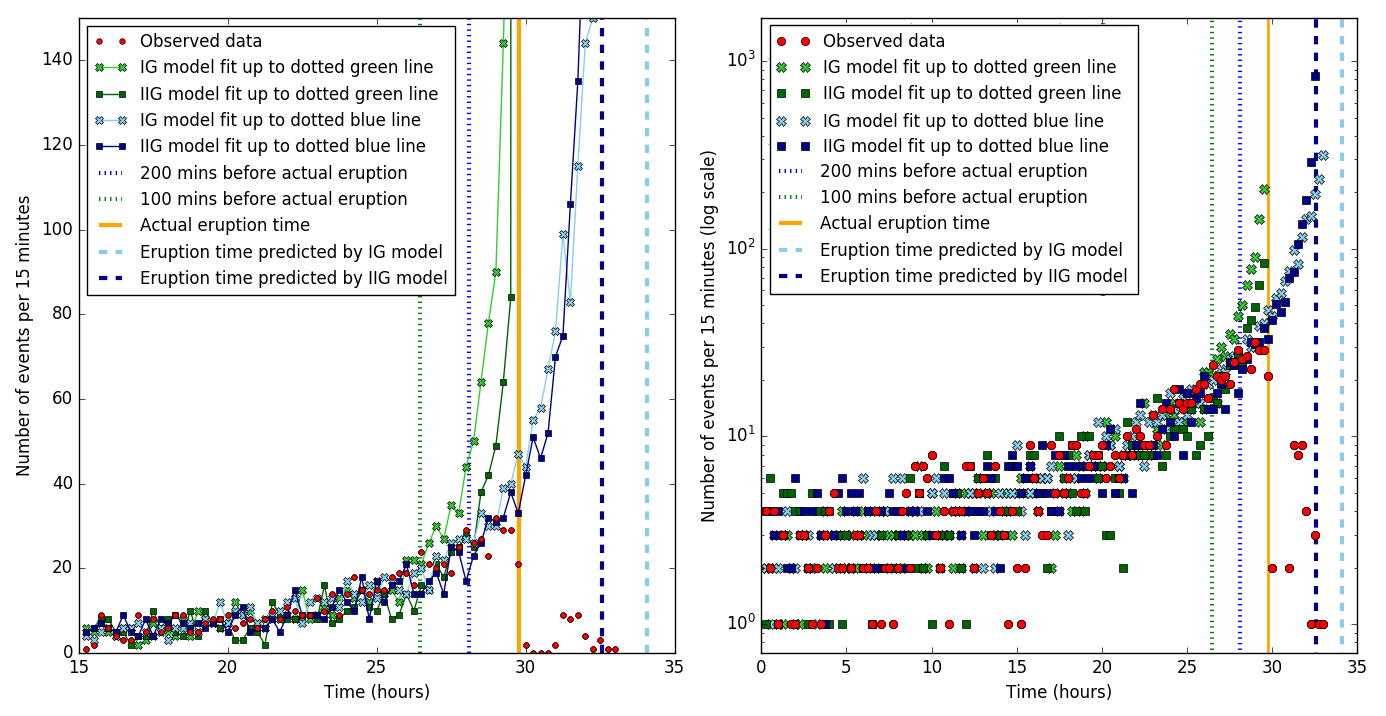}
\end{figure}

This behaviour can be explained by earthquake waveforms merging near eruption time \cite{bell2}. As spikes begin to occur at a higher rate, it becomes increasingly difficult to distinguish one event from another, resulting in an underestimation of the actual number of events occurring. Although advanced picking methods might minimise this effect, the very high rates and long coda suggest it will persist to some degree. Further work might include a template cross-correlation picking method to better identify events when mean ISI reduces below the event coda duration. This inaccuracy causes the estimates to become increasingly less reliable, as with all of the models we would expect more spikes to be occurring close to the eruption than are being observed. It also means that it is difficult to identify any real deviations from the power-law increase in rate towards eruption predicted by the FFM. It can be seen from Figure \ref{fig:fc_observed_IG_IP_IW} that the cut-off point for the dataset used in estimating the parameters (shown by the pink dashed line) appears appropriately chosen. 

To demonstrate the effect of setting the cut-off point closer to the eruption time, we use an alternative cut-off of 100 minutes prior to eruption. Figure \ref{fig:IG_IIG_incompleteness} demonstrates the difference between the event rate of the observed data (red points) and the expected event rates based on the IG and IIG models. The green models are fit up to 200 minutes before eruption, with $t_f$ set to the known actual eruption time. The blue models were fit using data up to 100 minutes before eruption. The plots show event rates calculated using a sample of data simulated from each of the fitted models.

The blue IIG model is closer to the observed data for longer, up to around the 29 hour mark, which is consistent with the findings of the Q-Q and K-S plot analysis. However, it is known that some of the events occurring near the eruption are missing, but the IG and IIG models try to provide a best fit for the data ignoring this fact. Thus, they fit the observed data for too long, overestimating $t_f$, when in reality they should be predicting a higher event rate earlier on. 

Using data up to the dotted green line (200 minutes before eruption) and setting $t_f$ to the known value, the light green IG model starts deviating from the real data earlier. The green IIG model adheres to the observed data longer. This again supports the conclusion that while the IIG model fits the data more closely, it may be more significantly affected by the fact that some of the data is known to be missing, therefore providing less accurate forecasts for eruption time. 

\subsection{Extent of incompleteness}

\begin{figure}[!htbp]
  \caption{Fitted IP models demonstrating effect of data incompleteness. The models are fit using different cut-offs, of 100 (orange dashed line) and 200 (pink dashed line) minutes before the eruption. Red dots represent the observed data. \label{fig:IP_cumulative}}
  \centering
    \includegraphics[width=0.5\textwidth]{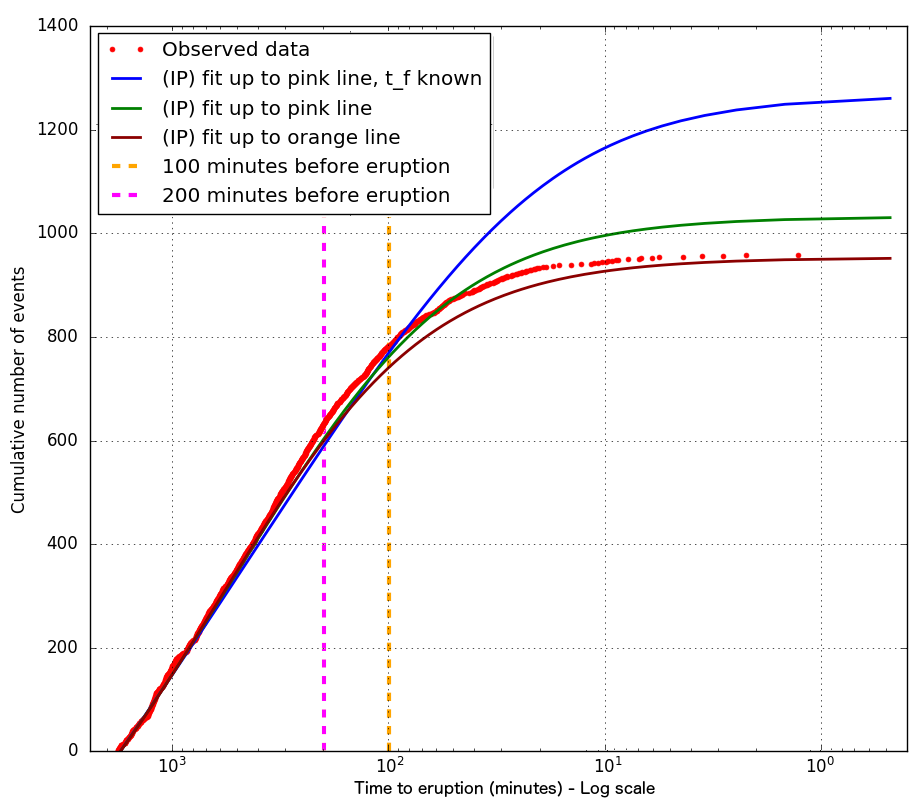}
\end{figure}

Figure \ref{fig:IP_cumulative} shows the cumulative number of events for the observed data (as red points), and the expected cumulative number of events for three IP models (as lines). 

The blue line shows the IP model fit up to 200 minutes before the eruption, with $t_f$ set to the actual value. This demonstrates approximately what we would expect to see if the merging of events were not an issue. The difference between the red line and the green line shows the difference that using data up to 100 mins vs. 200 mins before eruption time makes to the parameter estimates. The green line (IP model fit up to 200 mins before eruption) predicts more events than the dark red line (IP model fit up to 100 mins), however it is still significantly lower than the blue line, suggesting that the effect of merging events cannot be completely removed just by using data further away from the eruption. This suggests other diagnostics and measurements may be required to quantify and remove this effect. It is also possible that the lack of fit close to eruption is (partly) due to an underlying change in the process, or increasingly complexity not accounted for by the model.

\subsection{Effect of periodicity on estimation of $t_f$}

Bell et al. (2018) note that the variance of the posterior distribution of $t_f$ is lower for data simulated with a higher value of periodicity. 

\begin{figure}[!h]
  \caption{Effect of different values of $\alpha$ on forecast of $t_f$. Left panel (a) shows means (solid lines) and 95\% HPDIs (dashed lines) for forecasts of $t_f$ using simulated data with $\alpha$ of 1, 2.05, 5, 10 and 20. Right panel (b) shows width of 95\% HPDIs over time to eruption.\label{fig:alpha_v_tf}}
  \centering
    \includegraphics[width=\textwidth]{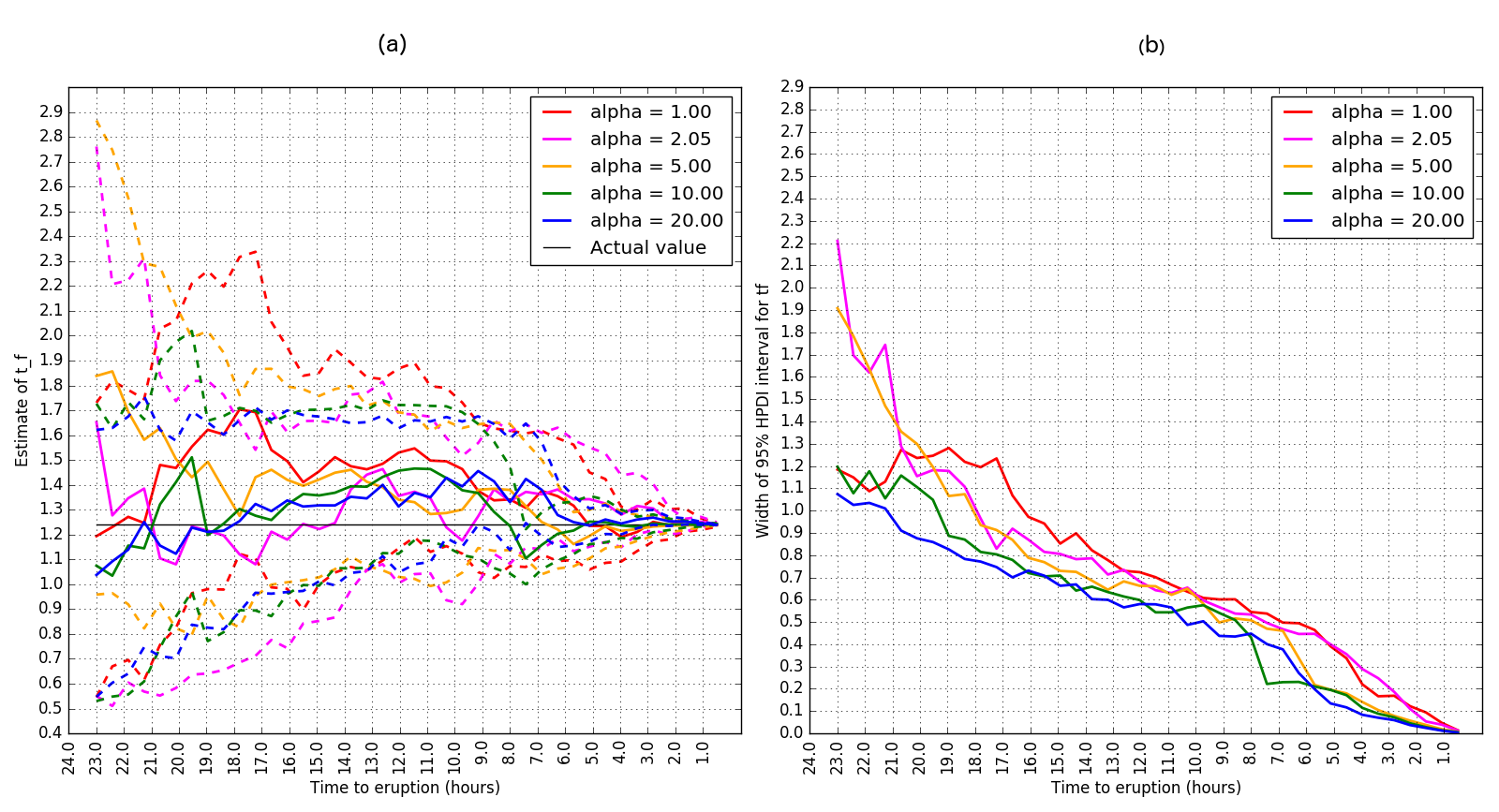}
\end{figure}

First, to explore the relationship between $\alpha$ and the forecast of $t_f$, five datasets were simulated from the IG model (with $t_f$ retrospectively set to the known value of 1.241 days) with five different values of $\alpha$ (1, 2, 5, 10 and 20). Then, hindcast forecasts were constructed similarly to the previous section, by re-fitting the IG model to increasing subsets of this simulated data. The resulting evolution of the forecast values of $t_f$ is shown in Figure \ref{fig:alpha_v_tf}(a). Figure \ref{fig:alpha_v_tf}(b) shows the widths of the resulting 95\% HPDIs.

It can be seen that as $\alpha$ increases, the HPDI width for $t_f$ reduces. This can be explained by the fact that larger $\alpha$ results in the mode of the distribution moving to the right; smaller $\alpha$ gives a mode very close to 0 with a larger right tail. Larger $\alpha$ results in a more narrow range of likely ISI values. Smaller $\alpha$ results in a wider range of likely ISI values. The reduction in the width of the HPDI for estimates of the parameters with larger $\alpha$ is due to the reduced range of random variation of ISIs. This, therefore, suggests that for forecasting purposes, the use of the IP model (with $\alpha = 1$) is justified, it will just result in a wider HPDI (i.e. higher uncertainty) for $t_f$. 

Further, for the observed dataset, the forecasting power of the IG and IP models was noted to be very similar, however as the periodicity of the data increases significantly, the IG model results in much lower uncertainty around the estimate of $t_f$ than the IP model. This is as expected, as the IG model allows for adjusting the $\alpha$ parameter which relates to periodicity, while for the IP model periodicity is fixed at 1.
\begin{figure}[!h]
  \caption{Evolution of $t_f$ for simulated data, with $\alpha$ set to 2, 10, 20 and 100. Dashed lines show 95\% HPDI boundaries, solid lines show posterior means for $t_f$.  \label{fig:simdata_forecast}}
  \centering
    \includegraphics[width=0.4\textwidth]{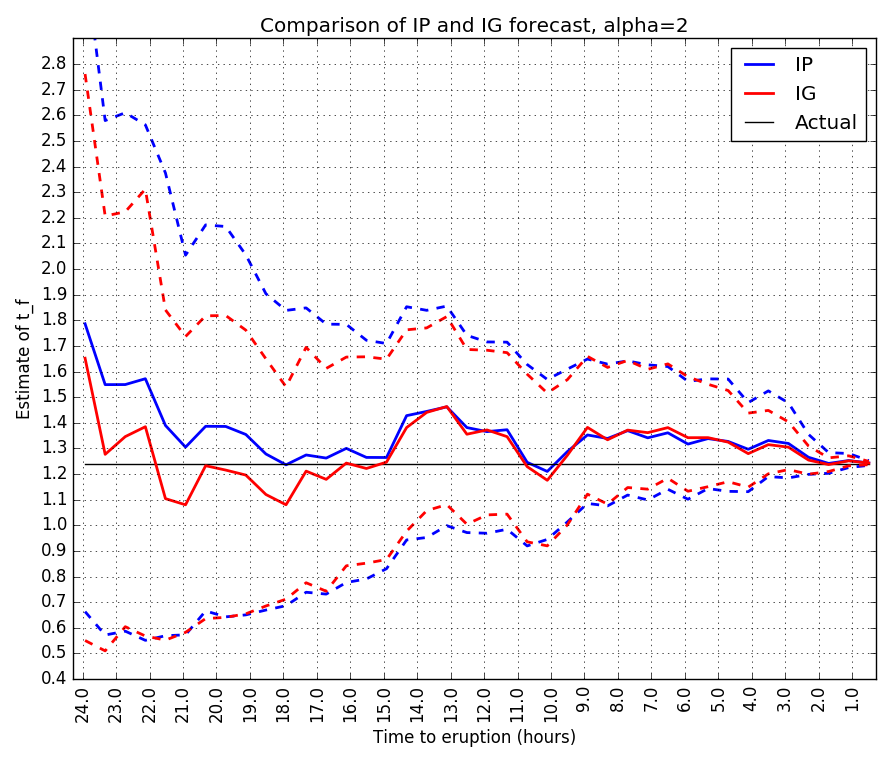}
    \includegraphics[width=0.4\textwidth]{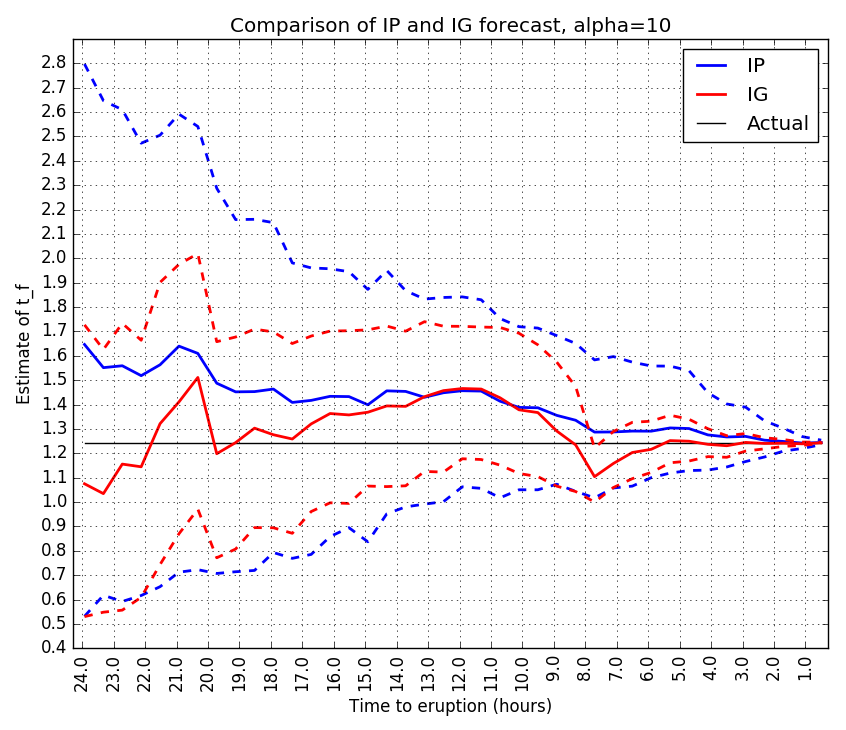}
    \includegraphics[width=0.4\textwidth]{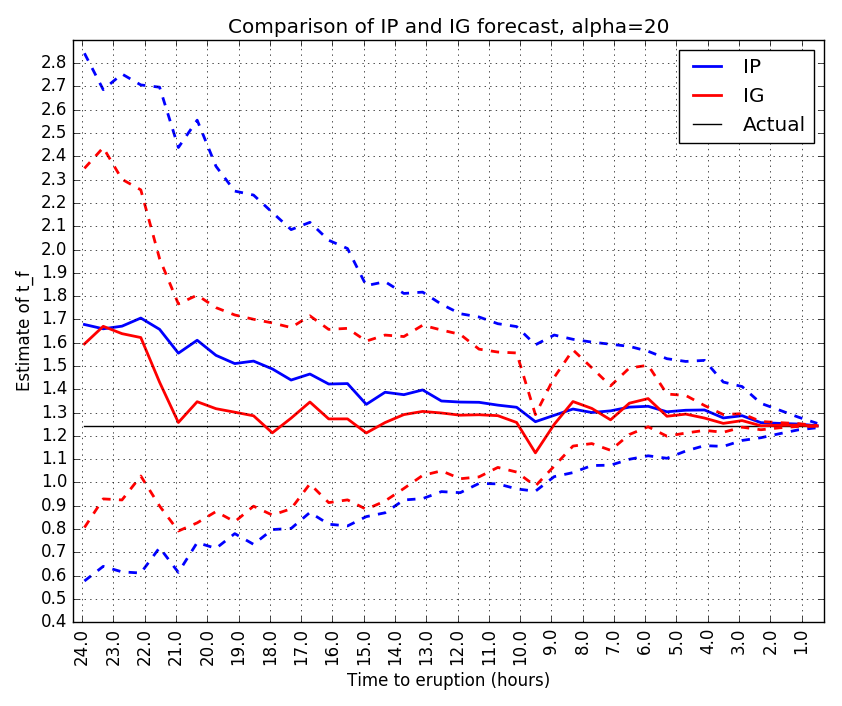}
    \includegraphics[width=0.4\textwidth]{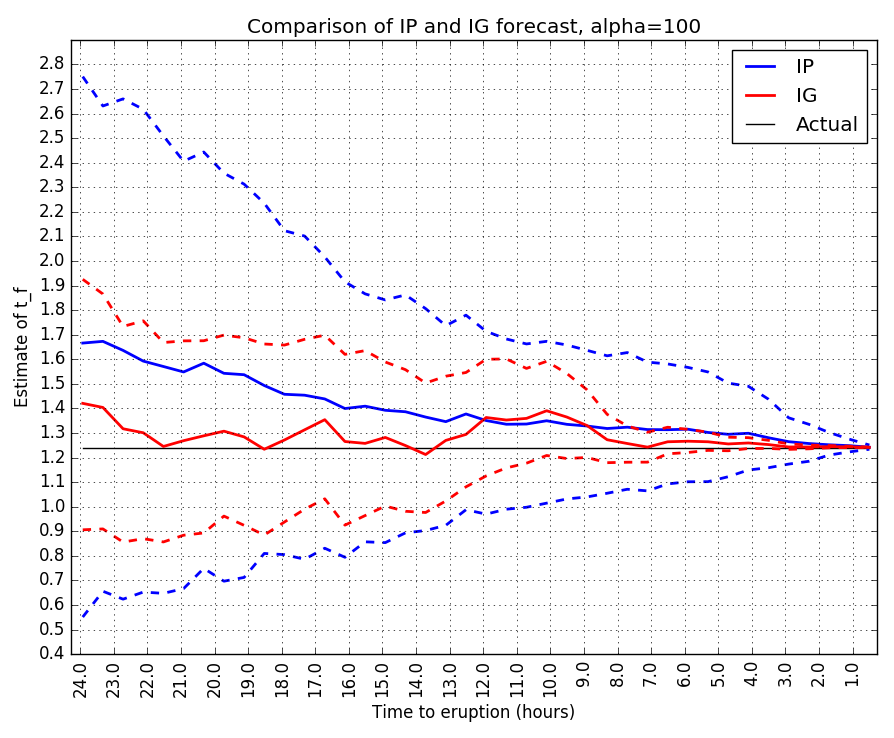}
\end{figure}

To demonstrate this difference between the IG and IP models, four datasets were simulated from the IG model with $\alpha$ equal to 2, 10, 20 and 100, to produce data that could be observed if the process followed the IG model exactly, with different periodicities. Then the IP and IG models were re-fit to this simulated data, to look at how the parameter estimates evolve over time. The resulting forecast plots in Figure \ref{fig:simdata_forecast} demonstrate that with increasing values of $\alpha$, the HPDIs for $t_f$ get narrower with the IG model with time, but stay roughly the same with the IP model. This suggests that the IP model may not be suitable for forecasting if the IG model is appropriate but the data is highly periodic.

\subsection{Effect of prior sensitivity on forecasts}

In fitting each of the models, we chose prior distributions for each of the power-law parameters $k$, $t_f$ and $p$ described in Section \ref{FFM}, to encapsulate information known about them before considering the given data. In this section, we explore the effect that changing these distributions would have on the results of the analysis.

For constructing prior distributions, there is no general rule which would help narrow the range for the more likely values of $k$ or $t_f$. However, there are prior beliefs regarding the power law coefficient $p$ which can be used. In the literature, the likely range of $p$ is generally given as 0.5 to 1.9 \cite{ogata1999,utsu1995,wiemer1999}, with a value near 1 being most likely. 

This information is reflected in the lognormal prior used:
\[ p \sim \text{logN}(0.1,0.25) \]
as most of the density lies between 0.5 and 2, with a mode around 1. To determine how sensitive the forecasting is to the specification of prior on $p$, a flat prior $U(0.5,2)$ was fit, as this constrains the range to the likely values without specifying a preference. 

Using the uniform prior results in slightly higher mean estimates of $t_f$ throughout the time period, around 2.4 hours greater than with the lognormal prior, up until around 3 hours before eruption when the two estimates converge to the true value. The 95\% HPDIs appear to be approximately the same throughout. At 200 minutes before eruption, both priors result in the same mean value (within 0.01), suggesting that the choice of prior would not significantly influence the resulting posterior parameter values for the models fit to the data. 

\section{Discussion}
 
The precursory LP earthquake sequence for the July 2013 eruption is empirically consistent with FFM, even though several aspects of the data are difficult to reconcile with a material failure model. It is possible that in this instance the LP earthquakes involved a localised material failure process, e.g. relating to a small asperity on the conduit margins. Alternatively, the sequence might be underpinned by a different physical process that results in apparently similar trends, e.g. accelerating ascent of the magma column controlled by rheological properties. It is not simple to distinguish between these models, particularly when considering only a single sequence. We hope that the methods we present here can allow improved statistical quantification of many pre-eruptive sequences, and additional forms of monitoring data (ground deformation, gas flux) when available, to begin to answer these questions.
 
Here we summarise our key findings, and discuss their implications for both retrospective data modelling and prospective forecasting. 
 
\subsection{Retrospective data modelling}
 
Retrospective analysis of pre-eruptive earthquake data is important for constraining the physical processes controlling the approach to eruption, and developing prior knowledge of model parameter values to improve future forecasting performance. From analysis of Q-Q and K-S plots, and simulated data, it is apparent that the IG model fits the July 2013 data very well. A small number of outliers (around 5\% of the data) was noted in the K-S plot, and correspond to spikes with long preceding ISIs. Some lack of fit was also found in the middle quantiles of the K-S plot, however this only slightly breached the 95\% error bounds. The IIG model was found to provide a slightly better overall fit than the IG model, and the IW and IP models a poorer fit. 
In practice, confidently distinguishing between the IG and IIG models in the presence of incomplete and noisy data, and a simplified underlying rate model, is not going to be possible without considering a greater number, and perhaps even better quality, datasets. The distributions of FFM parameter values across many pre-eruptive sequences, and different eruption styles and volcanoes, have not been comprehensively established. Likewise, the extent of applicability of the FFM, and the degree of quasi-periodic behaviour is largely unknown. These new methods provide a framework for such studies, and comparison between different volcanoes.

\subsection{Controls on forecasting performance}
Despite the better retrospective fit to the data, the IIG model was found to provide much less accurate forecasts of eruption time than the IG model. Although the IP model provides a very poor fit for all of the datasets, it provides eruption time forecasts which are very close to those of the IG model. This is explained by the fact that the IP model is a simplification of the IG model obtained by setting $\alpha=1$, and the finding that $\alpha$ and $t_f$ are relatively uncorrelated. Data with higher periodicity results in narrower credible intervals for $t_f$ for the IG model, but not the IP model. This explains why the IP model produces $t_f$ estimates with slightly wider 95\% HPDIs than the IG model.  
 
The earthquake catalogue is likely to miss up to 5\% of events, due to it occasionally being difficult to distinguish separate events from background activity, or two merging spikes appearing as one event. Analysis of ``€˜incomplete" simulated data with 5\% of the events uniformly removed over time resulted in the forecast of $t_f$ being consistently higher (i.e. predicting the eruption later). This should be expected, as having fewer events suggests that the eruption is further away in the future. However, the effect does not appear to be very large, with the forecasts differing by approximately 1-2 hours throughout, converging to the same value around 2.5 hours before eruption. 
 
Forecasts are significantly affected by the fact that near the eruption, it becomes increasingly difficult to distinguish individual earthquakes from each other in the seismic data. This causes fewer events to be recorded near the eruption than expected, causing inaccurate parameter estimates within around 3 hours of eruption. Excluding data within 200 minutes of the eruption when fitting the models lessens the negative effect on the parameter estimates, but does not remove it entirely, suggesting that other measurements and diagnostics need to be used to improve forecasting accuracy. It is also possible that the process changes close to eruption, resulting in a change in the event rate or the ISI distributions, or introducing complexity that is not appropriately accounted for by the point process models. 
 
\section{Conclusions}
The methodology and results presented here provide a framework for modelling and statistical analysis of accelerating rates of quasi-periodic seismicity before volcanic eruptions. We propose 4 candidate ISI distributions within a inhomogeneous point process model description of the FFM. Models based on the gamma and inverse Gaussian ISI distributions provide the best retrospective fits to the data. Models based on the gamma and exponential ISI distributions provide the best pseudo-prospective forecasts, though simulations show that with increased periodicity, the gamma variant will give narrower HDPIs. Outliers in the data can be explained by an approximate 5\% missed events within the catalogue, though real deviations from the simple underlying rate model are likely and cannot be excluded.
 
\section{Acknowledgements}

We thank the two anonymous reviewers and the editors for their helpful and constructive comments that have greatly improved the clarity of this paper.
 
\bibliography{literature_afb2}{}
\bibliographystyle{statsy}

\end{document}